\documentclass[10pt]{article}

\usepackage[a4paper,top=2.54cm,bottom=2.54cm,left=3.18cm,right=3.18cm]{geometry}

\usepackage{graphicx,color,xcolor,subcaption}
\usepackage{amsmath,amssymb,amsthm,mathtools,amsfonts,mathrsfs,bm}
\usepackage{algorithm,listings}
\usepackage{algpseudocode}
\usepackage{titlesec}
\usepackage{xr-hyper}
\usepackage{hyperref,url}
\usepackage{authblk}
\usepackage[version=4]{mhchem}

\usepackage[numbers]{natbib}
\newtheorem*{theorem}{THEOREM}

\title{Approximation Error of the Burst Approximation for a Stochastic Gene Expression Model}

\author{Yuntao Lu \thanks{Email: \href{mailto:yuntaolu22@m.fudan.edu.cn}{yuntaolu22@m.fudan.edu.cn}}}

\author{Yunxin Zhang \thanks{Email: \href{mailto:xyz@fudan.edu.cn}{xyz@fudan.edu.cn}}}

\affil{School of Mathematical Sciences, Fudan University,\\ Shanghai 200433, CHINA}

\date{\today}

\begin{document}

\maketitle
\begin{abstract}
Stochastic modeling of gene expression is a classic problem in theoretical biophysics, and the burst approximation is widely used to simplify gene expression models formulated via the chemical master equation. However, the approximation error has been investigated only for the simplest case. This article proposes and analyzes a general stochastic gene expression model with an arbitrary number of gene states, and quantifies the error introduced by the burst approximation. Using the standard binomial moment method, we derive recurrence relations for binomial moments in steady state. We develop an algorithm to numerically compute binomial moments in a hierarchical manner. In particular, explicit expressions for low-order moments are presented. Compared with surrogate models under the burst approximation, we conclude that the first-order moment of protein counts is preserved, whereas discrepancies generally arise in higher-order moments. By estimating the difference between two second-order moments using functional analysis, we evaluate the validity of the burst approximation. 
\end{abstract}

\tableofcontents
\section{Introduction}
Gene expression is one of the most fundamental processes in molecular biology, through which genetic information is converted into functional gene products. In most cases, gene expression involves two main stages: transcription and translation. During transcription, messenger RNA (mRNA) is synthesized from specific segments of DNA, whereas translation refers to the production of proteins using mRNA as a template. To quantitatively understand and predict this process, it is essential to propose appropriate mathematical models. Notably, intrinsic stochasticity is the key feature of gene expression. Such inherent stochasticity originates from random collisions among molecules, and becomes relatively significant due to low numbers of mRNA and protein molecules in most cells \cite{CentralDogmaSinglemolecule2011}. Randomized expression of genes can be beneficial for biological adaptation to changing environments \cite{NoiseGeneExpression2005}. However, how these fluctuations propagate within and across cells, and how they are ultimately regulated, remains an open question at the theoretical level.

The primary variables are the mRNA and protein copy numbers during gene expression, whose dynamics may be abstracted as a stochastic process. Note that deterministic approaches like reaction rate equation \cite{AppliedStochasticAnalysis2019} usually fail to describe this process because of pronounced stochasticity. More specifically, this stochastic process should be a continuous-time process with a discrete state space. The state space should be discrete because the mRNA and protein counts are nonnegative integers. The simplest class of stochastic processes satisfying both conditions consists of continuous-time Markov chains, whose Kolmogorov forward equation is commonly termed the chemical master equation \cite{AppliedStochasticAnalysis2019,StochasticProcessesPhysics2007} in chemical physics literature. In general, the chemical master equation is an infinite-dimensional coupled system of ordinary differential equations, and can be equivalently converted into a finite-dimensional partial differential equation system using the generating function method or the Laplace transform. When analyzing a dynamical system described by the chemical master equation, neither a universal treatment nor an effective general theory exists, and one usually devises techniques tailored to different problems. In particular, concise analytical results are rarely anticipated, and in many cases, different kinds of approximation techniques are applied \cite{ApproximationInferenceMethods2017}. The most widely used approximation method is the diffusion approximation \cite{ChemicalLangevinEquation2000}, which approximates the chemical master equation with a Markov-type stochastic differential equation \cite{BrownianMotionStochastic1998}, conventionally termed the chemical Langevin equation. This procedure approximates a jump process with a diffusion process. Therefore, the diffusion approximation may become inappropriate when the numbers of molecules in the system are extremely low, as the discreteness of the state space becomes significant. Despite the challenges, gene expression models based on the chemical master equation are worth in-depth analysis given their biological importance.

Proposing stochastic gene expression models is straightforward. The simplest two-stage model is analyzed in \cite{AnalyticalDistributionsStochastic2008}. In the two-stage model, mRNA molecules are continuously produced according to a Poisson process, and translation, hydrolysis of mRNA, and hydrolysis of protein all occur as memoryless single-step reactions. Using the generating function method and the method of characteristics for solving linear partial differential equations, the authors of \cite{AnalyticalDistributionsStochastic2008} obtain a formal analytical expression of the generating function. However, the expression of the generating function contains an incomplete gamma function that cannot be reduced, and an analytical expression for the protein copy number distribution is infeasible \cite{AnalyticalDistributionsStochastic2008}. For a long time, results on stochastic gene expression models are largely confined to low-order moments of several toy models \cite{ModelsStochasticGene2005}. In \cite{ExactDistributionsFull2017}, the binomial moment method \cite{BinomialMomentEquations2011,StochasticAnalysisComplex2012,MomentconvergenceMethodStochastic2016} is first applied to a full gene expression model with two gene states. Both recurrence relations for binomial moments and analytical expressions for low-order moments are provided \cite{ExactDistributionsFull2017}. Similar methodology is also used to study a full gene expression model with multiple active and inactive gene states \cite{InfluenceComplexPromoter2019}. Related work also includes \cite{StationarityInferenceMultistate2022}.

Meanwhile, an effective approximation technique, namely, the burst approximation \cite{ModelsStochasticGene2005,AnalyticalDistributionsStochastic2008}, is also available. The burst approximation builds upon the experimental conclusion that mRNAs decay substantially faster than proteins in most cells \cite{AnalyticalDistributionsStochastic2008}. Under the burst approximation, the original gene expression models can be greatly simplified and analytical results become readily available \cite{ExactDistributionsStochastic2022,AnalyticalTimedependentDistributions2023,AnalyticalDistributionsStochastic2008}. Beyond their roles as approximate substitutes, stochastic gene expression models with translational bursting have independent theoretical value. Particularly, models with multiple gene states under the burst approximation have been studied using the binomial moment method \cite{ExactDistributionsStochastic2022}, and non-Markovian models can also be studied using techniques from queueing theory \cite{IntrinsicNoiseStochastic2011}. According to the equivalence between queueing systems and gene expression models established in \cite{SolvingStochasticGeneexpression2024}, models under the burst approximation can be interpreted as queueing systems with batch arrivals. This perspective may explain the substantial analytical simplifications introduced by the burst approximation, and clarify the theoretical significance of related research. Nevertheless, whether the errors introduced by the burst approximation are controllable has not received much attention. The validity of the burst approximation has only been evaluated in the simplest two-stage model \cite{AnalyticalDistributionsStochastic2008}, and validation for general models is the main motivation in this article.

Additionally, general numerical methods exist for reaction systems described by the chemical master equation. The stochastic simulation algorithm (SSA) \cite{GeneralMethodNumerically1976} is a numerical method for generating the sample paths of a given continuous-time Markov chain, one realization at a time. Multiple variants of the SSA are also available \cite{StochasticSimulationChemical2007}. In this article, we employ the Python package \texttt{GillesPy2} \cite{GillesPy2BiochemicalModeling2023} to perform the SSA. The finite state projection algorithm (FSP) \cite{FiniteStateProjection2006} is also a widely used numerical method, which truncates the chemical master equation and reduces the problem to numerically solving an ordinary differential equation system with constant coefficients. In general, numerical methods including the SSA and the FSP are computationally expensive and quickly become infeasible for large systems. Moreover, the aforementioned numerical methods cannot be used to derive further theoretical results. As a result, theoretically analyzing complete stochastic gene expression models is still important. 

In this article, we analyze a general complete gene expression model with multiple gene states, and estimate the error introduced by the burst approximation. 
Firstly, similar to \cite{ExactDistributionsFull2017,InfluenceComplexPromoter2019}, we apply the high-dimensional binomial moment method to derive recurrence relations for binomial moments of mRNA and protein copy number, thereby yielding exact analytical expressions for binomial moments of arbitrary order. This part is a straightforward generalization of the results in \cite{ExactDistributionsFull2017,InfluenceComplexPromoter2019}. Next, we design an algorithm that numerically computes, with high accuracy, all binomial moments in a hierarchical way. When binomial moments are obtained, the joint probability mass function of mRNA and protein copy numbers can be constructed directly. In particular, we present explicit expressions for low-order moments. Compared with models under the burst approximation, we conclude that the first-order binomial moment remains exact, whereas the second-order binomial moment generally differs. An upper bound is given for the difference between the second-order binomial moments obtained from two models, mainly using techniques from functional analysis. Finally, we present the analytical expression of the multi-variable generating function by solving a partial differential equation system.

The article is structured as follows. In \autoref{sec2}, a general stochastic gene expression model is introduced, in which the gene has arbitrarily many states and can transition freely among them. In \autoref{sec3}, the chemical master equation describing the dynamics is presented, and then converted to a partial differential equation system using the standard generating function method. Using high-dimensional binomial moment method, an ordinary differential equation hierarchy governing the evolution of binomial moments of mRNA and protein copy number is derived. We present the concise recurrence relations in steady state and design an algorithm for numerical computation. These results are presented in \autoref{sec4}. In \autoref{sec5}, we provide an equality to reconstruct the probability mass function from binomial moments. Analytical expressions for low-order cumulants of the number of mRNA and protein molecules are given in \autoref{sec6}, and are compared with those from models under burst approximation in \autoref{sec7}.

\section{Stochastic Gene Expression Model}
\subsection{Model Description}\label{sec2}
We now introduce a general stochastic model of gene expression, formulated as a chemical reaction system schemed in \eqref{Reaction1}. In this model, the gene we consider is assumed to have $N$ different states, namely, $\mathcal{S}_i\;(1\leq i \leq N)$, and the gene transitions arbitrarily among these states in a Markovian manner. In each state, mRNA molecules are transcribed at a different rate and the gene state is allowed to switch upon transcription. Transcribed mRNA molecules are confronted with competing reaction pathways of hydrolysis and translation. Hydrolysis of mRNA molecules occurs at a constant rate, independent of the state of the gene or the number of protein molecules. A specific type of protein molecules is translated from living mRNA molecules at constant rate, and also undergo hydrolysis once produced. Based on the general theory of stochastic chemical reaction kinetics \cite{AppliedStochasticAnalysis2019}, \eqref{Reaction1} defines a continuous-time Markov chain $(S(t), M_1(t), M_2(t))_{t\geq 0}$ with state space $\{1,2,\cdots,N\}\times\mathbb{N}\times\mathbb{N}$, where $S(t)$, $M_1(t)$ and $M_2(t)$ denote the state of the gene, the number of mRNA molecules in the system, and the number of protein molecules in the system. 

\begin{equation}\label{Reaction1}
\begin{aligned}
    &\ce{\mathcal{S}_i ->[$a_{i,j}$] \mathcal{S}_j}\;\;(i\neq j,\;\;1\leq i,j\leq N)\\
    &\ce{\mathcal{S}_i ->[$b_{i,j}$] \mathcal{S}_j + \textbf{mRNA}}\;\;(1\leq i,j\leq N)\\
    &\ce{\textbf{mRNA} ->[$u$] $\emptyset$}\\
    &\ce{\textbf{mRNA} ->[$v$] \textbf{mRNA} + \textbf{Protein}}\\
    &\ce{\textbf{Protein} ->[$\delta$] $\emptyset$}\\
\end{aligned}
\end{equation}

Denote by $a_{i,j}\;(i\neq j,\;1\leq i,j\leq N)$ the transition rates among the states of gene without transcription, and $b_{i,j}\;(1\leq i,j\leq N)$ the transition rates with production of one mRNA molecule. Let $v$, $u$, and $\delta$ denote the translation rate, the mRNA degradation rate, and the protein degradation rate, respectively. Define $a_{i,i}:=-\sum_{\substack{k=1\\k\neq i}}^Na_{i,k}-\sum_{k=1}^Nb_{i,k}$, $D_0:=(a_{i,j})_{N\times N}$, $D_1:=(b_{i,j})_{N\times N}$, and $D:=D_0 +D_1$.

\subsection{Chemical Master Equation}\label{sec3}
Assume that there are no mRNA or protein molecules in the system at time $0$ and that the gene is in a given state, say, $S_i\;(1\leq i\leq N)$. Denote by $\mathbb{P}_{i,j}(m,n;t)\;(1\leq j\leq N, m\in\mathbb{N},n\in\mathbb{N}, t\geq 0)$ the probability that at time $t$ there are $m$ mRNA molecules, $n$ protein molecules and the gene is in state $S_j$. Define $\mathbb{P}(m,n;t)$ to be a $N\times N$ matrix whose $(i,j)$-th element is $\mathbb{P}_{i,j}(m,n;t)$. The chemical master equation of \eqref{Reaction1} is
\begin{equation}\label{CME1}
\begin{aligned}
\frac{\partial}{\partial t}\mathbb{P}_{i,j}(m,n;t)&=\sum_{\substack{s=1\\s\neq j}}^Na_{s,j}\mathbb{P}_{i,s}(m,n;t)+\sum_{s=1}^Nb_{s,j}\mathbb{P}_{i,s}(m-1,n;t)\\
&+mv\mathbb{P}_{i,j}(m,n-1;t)+(m+1)u\mathbb{P}_{i,j}(m+1,n;t)\\&+(n+1)\delta\mathbb{P}_{i,j}(m,n+1;t)-(mu+n\delta+mv)\mathbb{P}_{i,j}(m,n;t)\\&-(\sum_{\substack{s=1\\s\neq j}}^Na_{j,s}+\sum_{s=1}^Nb_{j,s})\mathbb{P}_{i,j}(m,n;t).
\end{aligned}
\end{equation}

Using the generating function method, or equivalently, the Laplace transform, we define the matrix-form generating function of $\mathbb{P}(m,n,t)$ as 
\begin{equation}\label{generating1}
\begin{aligned}
\mathcal{G}(z,w;t):=\sum_{m=0}^\infty\sum_{n=0}^\infty z^mw^n\mathbb{P}(m,n;t),\;\;z,w\in\mathbb{R},\;\lvert z \rvert \leq 1, \; \lvert w \rvert \leq 1.
\end{aligned}
\end{equation}
The infinite-dimensional ordinary differential equation system \eqref{CME1} can then be transformed into a finite-dimensional partial differential equation system, namely, 
\begin{equation}\label{partial1}
\begin{aligned}
\frac{\partial}{\partial t}\mathcal{G}(z,w;t)=&\mathcal{G}(z,w;t)D_0+z\mathcal{G}(z,w;t)D_1\\&+\left[u(1-z)+vz(w-1)\right]\frac{\partial}{\partial z}\mathcal{G}(z,w;t)+\delta(1-w)\frac{\partial}{\partial w}\mathcal{G}(z,w;t).
\end{aligned}
\end{equation}
Detailed derivation can be found in the Supplementary Material.

Remarkably, by setting $w=1$, we focus on the marginal distribution of mRNA copy number, and \eqref{partial1} reduces to
\begin{equation}\label{partial2}
\begin{aligned}
\frac{\partial}{\partial t}\mathcal{G}(z,1;t)=&\mathcal{G}(z,1;t)D_0+z\mathcal{G}(z,1;t)D_1+u(1-z)\frac{\partial}{\partial z}\mathcal{G}(z,1;t),
\end{aligned}
\end{equation}
which is exactly the partial differential equation system that appeared in our previous work \cite{StochasticKineticsMRNA2025}. This can be intuitively understood since the fluctuations of the protein copy number do not contribute to the fluctuation at mRNA level, while changes in mRNA copy number do affect fluctuations at protein level. This phenomenon is widely known as dynamic disorder \cite{SinglemoleculeApproachDispersed2002}. Additionally, the probability distribution of the protein copy number cannot be treated marginally; instead, the joint probability distribution needs to be considered.

Actually, we also construct the analytical solution to the partial differential equation system \eqref{partial2} using the method of characteristics, which can be found in the Supplementary Material. We note that further analysis based on this analytical solution requires additional work and is not pursued in this article. 

\section{Reconstruct Probability Mass Function from Binomial Moments}
\subsection{Two-dimensional Binomial Moment Method}\label{sec4}
In this and the following section, we aim to establish an efficient numerical scheme for computing the steady-state joint distribution of mRNA and protein copy number. Our approach builds on the standard binomial moment method \cite{BinomialMomentEquations2011,StochasticAnalysisComplex2012,MomentconvergenceMethodStochastic2016}, which analyzes the moments of a probability distribution and subsequently reconstructs the corresponding probability mass function. In fact, the binomial moment method can be used to study any first-order reaction system. We first obtain the binomial moments in this section.

Define the binomial moment of the joint probability distribution as
\begin{equation}\label{BM}
\begin{aligned}
    \mathcal{B}_{p,q}(t):=\sum_{m=p}^\infty\sum_{n=q}^\infty \binom{m}{p}\binom{n}{q}\mathbb{P}(m,n;t),\; p\in\mathbb{N},\; q\in\mathbb{N},
\end{aligned}
\end{equation}
where $\binom{m}{p}$ denotes the combinatorial coefficient.

Binomial moments are linear combinations of moments defined by $\langle M_1(t)^pM_2(t)^q\rangle$\newline$:=\sum_{m=0}^\infty\sum_{n=0}^\infty m^pn^q\mathbb{P}(m,n;t)$ \cite{CourseProbabilityTheory2000}, therefore binomial moments exist if and only if moments of arbitrary order exist for each element in $\mathbb{P}(m,n;t)$. The name ``binomial'' comes from combinatorial coefficients in \eqref{BM}. Additionally, denote the stationary binomial moments by $\mathcal{B}_{p,q}:=\lim_{t\rightarrow\infty}\mathcal{B}_{p,q}(t)$ and the stationary matrix-valued probability mass function by $\mathbb{P}(m,n):=\lim_{t\rightarrow\infty}\mathbb{P}(m,n;t)$.

According to the definition \eqref{BM}, we have 
\begin{equation}\label{related}
\begin{aligned}
    \mathcal{G}(z,w;t)&\equiv\sum_{m=0}^\infty\sum_{n=0}^\infty z^mw^n\mathbb{P}(m,n;t)
    =\sum_{p=0}^\infty\sum_{q=0}^\infty s^pr^q\mathcal{B}_{p,q}(t),
\end{aligned}
\end{equation}
where $s:=z-1$ and $r:=w-1$. Substituting \eqref{related} into \eqref{partial1}, we may obtain an ordinary differential equation hierarchy for $\mathcal{B}_{p,q}(t)$. Setting the time-derivatives to zero, we get a linear system for binomial moments at steady state. Specifically, we have 
\begin{equation}\label{BM_hierachy}
\begin{aligned}
   \mathcal{B}_{p,q}\left(D-up\bm{I}_N-\delta q\bm{I}_N\right)+v(p+1)\mathcal{B}_{p+1,q-1}+vp\mathcal{B}_{p,q-1}+ \mathcal{B}_{p-1,q}D_1=0,\;p\in\mathbb{N},\;q\in\mathbb{N}.
\end{aligned}
\end{equation}
Note that $\mathcal{B}_{0,0}=\bm{I}_N$ by definition and $ \mathcal{B}_{p,q}$ is taken as $\bm{0}_{N\times N}$ if $p<0$ or $q<0$. See Supplementary Material for the detailed derivation. \eqref{BM_hierachy} agrees with the results in \cite{ExactDistributionsFull2017,InfluenceComplexPromoter2019}.

\begin{figure}[h!]
    \centering
    \includegraphics[width=0.8\linewidth]{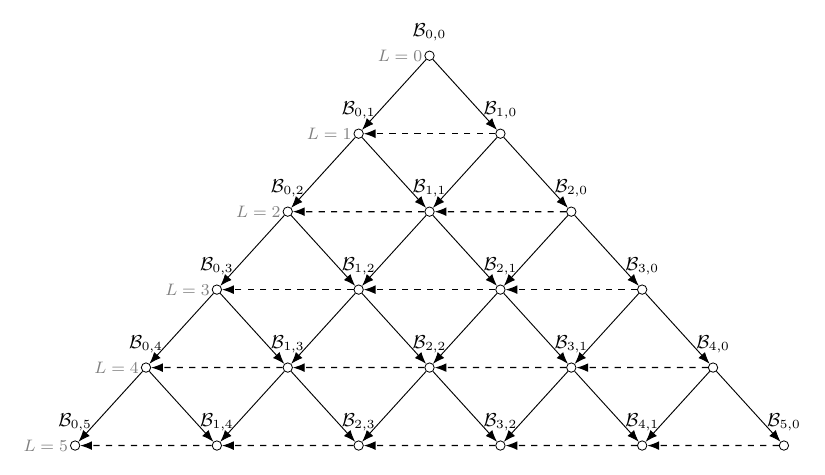}
    \caption{\textbf{Hierarchy of Binomial Moments}: This image illustrates the hierarchical structure for calculating binomial moments up to desired orders. The layers are organized from $L=0$ at the top to $L=5$ at the bottom. Within each layer, binomial moments are arranged from left to right in order of increasing $p$ (and correspondingly decreasing $q$). The calculation proceeds from lower to higher layers. Within each layer, binomial moments are derived sequentially from right to left. The arrows indicate the dependency flow, showing how binomial moments in higher layers depend on values calculated earlier in the same or preceding layers.}
    \label{hierarchy}
\end{figure}

Arbitrary-order binomial moments can be obtained in a hierarchical manner using \eqref{BM_hierachy}. Define the layer, or order, of a binomial moment $\mathcal{B}_{p,q}$ as $L:=p+q$. All binomial moments are assembled according to their layers, proceeding from lower to higher layers. Within each layer, binomial moments are ordered by increasing $p$ and derived sequentially from right to left. This overall procedure is illustrated in \autoref{hierarchy}, and the detailed algorithmic process is presented in Algorithm S.1. 
 
Recall that $D$ is the generator of the Markov chain $(S(t))_{t\geq 0}$. Therefore, $\gamma\bm{I}_N-D$ is a strictly diagonally dominant matrix for any $\gamma>0$, and is nonsingular by L\'{e}vy-Desplanques theorem \cite{MatrixAnalysis2012}. Additionally, strict diagonal dominance guarantees the numerical stability while performing a LU factorization \cite{MatrixComputations2013}. Thus, the numerical scheme Algorithm S.1 yields binomial moments with high accuracy. 

Note that analytical expressions of the binomial moments of mRNA copy number, namely, $\mathcal{B}_{p,0}\;(p\in\mathbb{N})$, agree with existing results \cite{MultimodalityFlexibilityStochastic2013,StochasticKineticsMRNA2025,PromotermediatedTranscriptionalDynamics2014}. However, the derivation here is primarily formal in comparison with our earlier work \cite{StochasticKineticsMRNA2025}, in which the time-dependent binomial moments are derived based on the analytical solution to the chemical master equation, and the stationary expressions are obtained by taking temporal limit. 

\subsection{Reconstruct Probability Mass Function}\label{sec5}
Under certain conditions, all moments can uniquely determine the underlying probability distribution \cite{CourseProbabilityTheory2000}. Particularly, the binomial moments may determine the probability mass function through a compact identity. According to \eqref{related}, stationary probability mass function $\mathbb{P}(m,n)$ can be reconstructed from binomial moments according to \cite{BinomialMomentEquations2011,StochasticAnalysisComplex2012,MomentconvergenceMethodStochastic2016}
\begin{equation}\label{reconstruct}
\begin{aligned}
    \mathbb{P}(m,n)=\sum_{p=m}^\infty\sum_{q=n}^\infty(-1)^{p+q+m+n}\binom{p}{m}\binom{q}{n}\mathcal{B}_{p,q}.
\end{aligned}
\end{equation}
See Supplementary Material for the detailed derivation.

We note that, in practice, computation based on \eqref{reconstruct} suffers from severe floating-point error if not implemented properly. 
This is because for large $p$ and $q$, multiplying relatively small binomial moment by extremely large combinatorial coefficients can introduce significant floating-point errors.
This is essentially the same numerical issue encountered and analyzed in \cite{ExactlySolvableModels2020,StochasticKineticsMRNA2025}. In the accompanying codes, we take the approach in \cite{StochasticKineticsMRNA2025}, carrying out computations in the logarithmic domain and exponentiating the final results.

\begin{figure}[t!]
    \centering
    \includegraphics[width=1\linewidth]{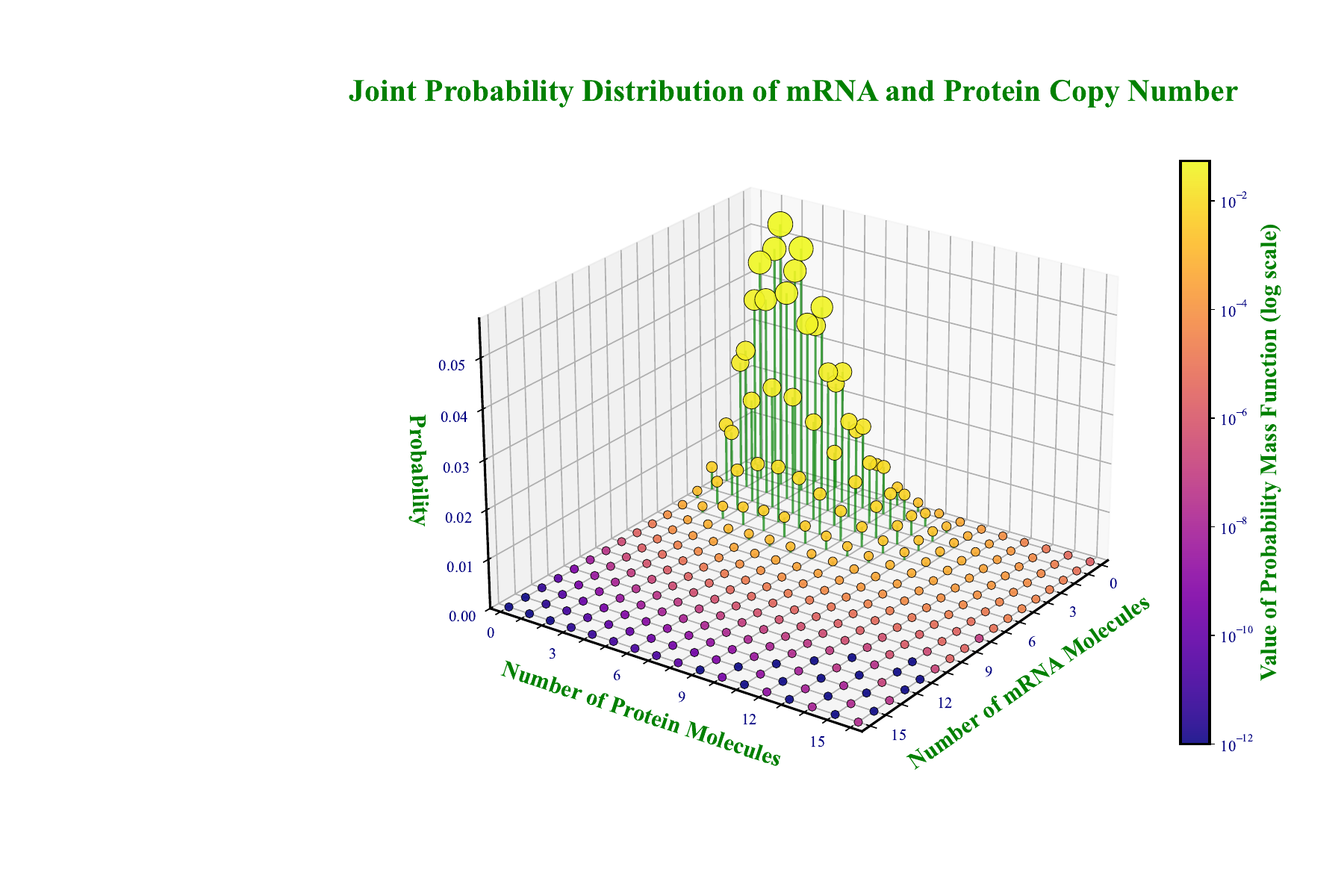}
    \includegraphics[width=1\linewidth]{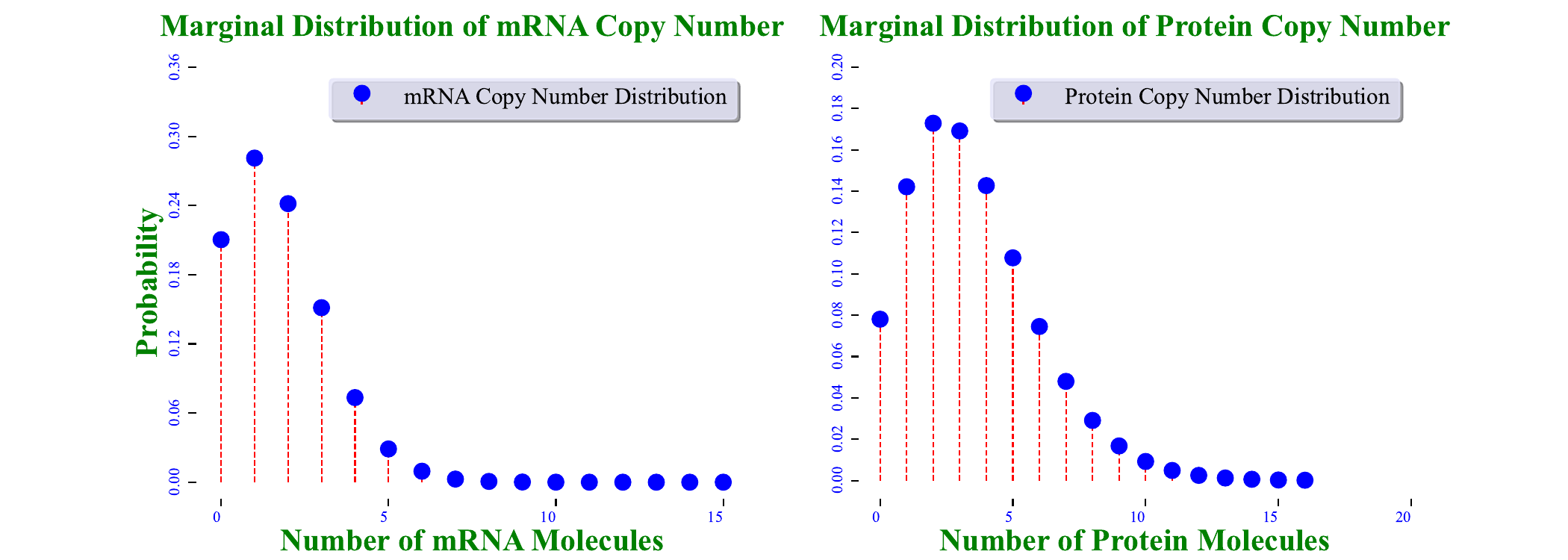}
    \caption{\textbf{Probability Distribution of mRNA and Protein Copy Number obtained using Analytical Results}: This stem plot illustrates the probability mass function obtained by first computing binomial moments according to Algorithm S.1, and then reconstructing via \eqref{reconstruct}. In this example, parameters in the model \eqref{CME1} are $D_0=$ \scalebox{0.5}{$\begin{pmatrix}
    -2.02 & 0.01 & 0.01 \\
    0.1  & -7.2 & 0.1  \\
    0    & 0.01 & -6.01
\end{pmatrix}$}, $D_1 =$ \scalebox{0.5}{$\begin{pmatrix}
    1 & 0 & 1 \\
    1 & 5 & 1 \\
    0 & 1 & 5
\end{pmatrix}$}, $u=3$, $v=2$, and $\delta=1$. The largest layer in Algorithm S.1 is $L_{\text{max}}=300$. $\mathbb{P}(m,n)$ is computed for $0\leq m,n\leq 16$. The infinite series \eqref{reconstruct} is truncated up to the largest layer $L_{\text{max}}$. In the stem plot in the top panel, we use the coarse-grained probability mass function $P(m,n)$ defined in \autoref{sec6}. The two stem plots in the bottom panel are the marginal distribution of mRNA and protein copy number, respectively. The marginal distributions are obtained directly from the joint probability distribution.}
    \label{our}
\end{figure}

\begin{figure}[t!]
    \centering
    \includegraphics[width=1\linewidth]{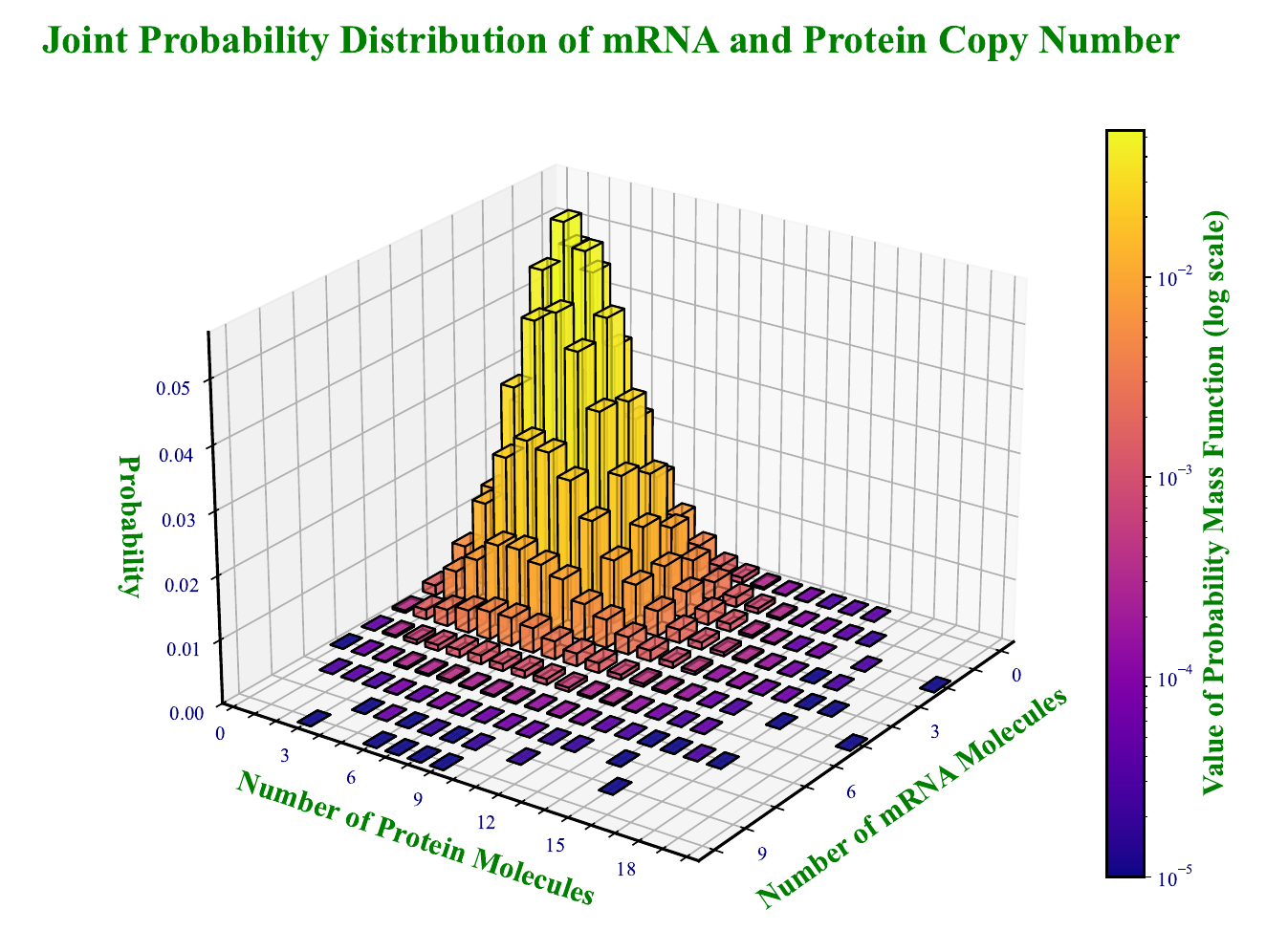}
    \includegraphics[width=1\linewidth]{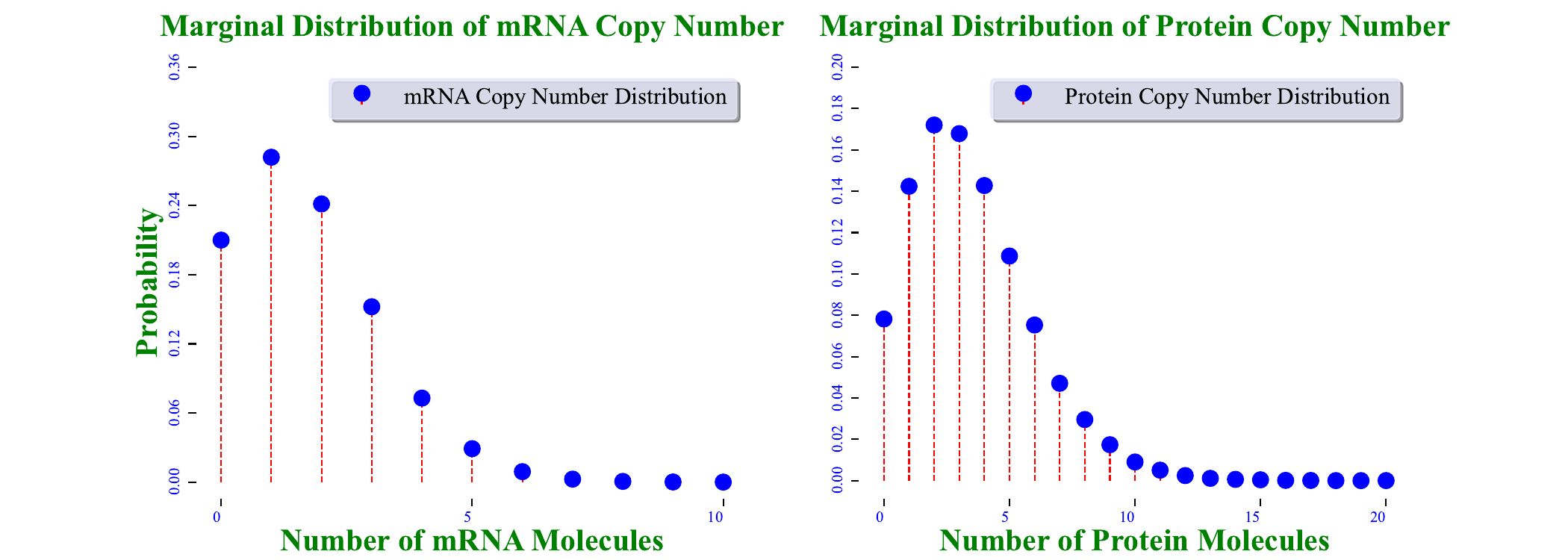}
    \caption{\textbf{Probability Distribution of mRNA and Protein Copy Number obtained through Stochastic Simulations}: The parameters in the model \eqref{CME1} are the same as those in \autoref{our}. The histogram in the upper panel is generated with $1\times10^6$ trajectories of SSA, all truncated at dimensionless time $t=50$. The initial condition is $S(0)=1$, $M_1(0)=0$, and $M_2(0)=0$. Python package \texttt{GillesPy2} is implemented with C++ solver. The two stem plots in the bottom panel are the marginal distribution of mRNA and protein copy number, respectively. The marginal distributions are obtained directly from the joint probability distribution.}
    \label{SSA}
\end{figure}

\begin{figure}[t!]
    \centering
    \includegraphics[width=1\linewidth]{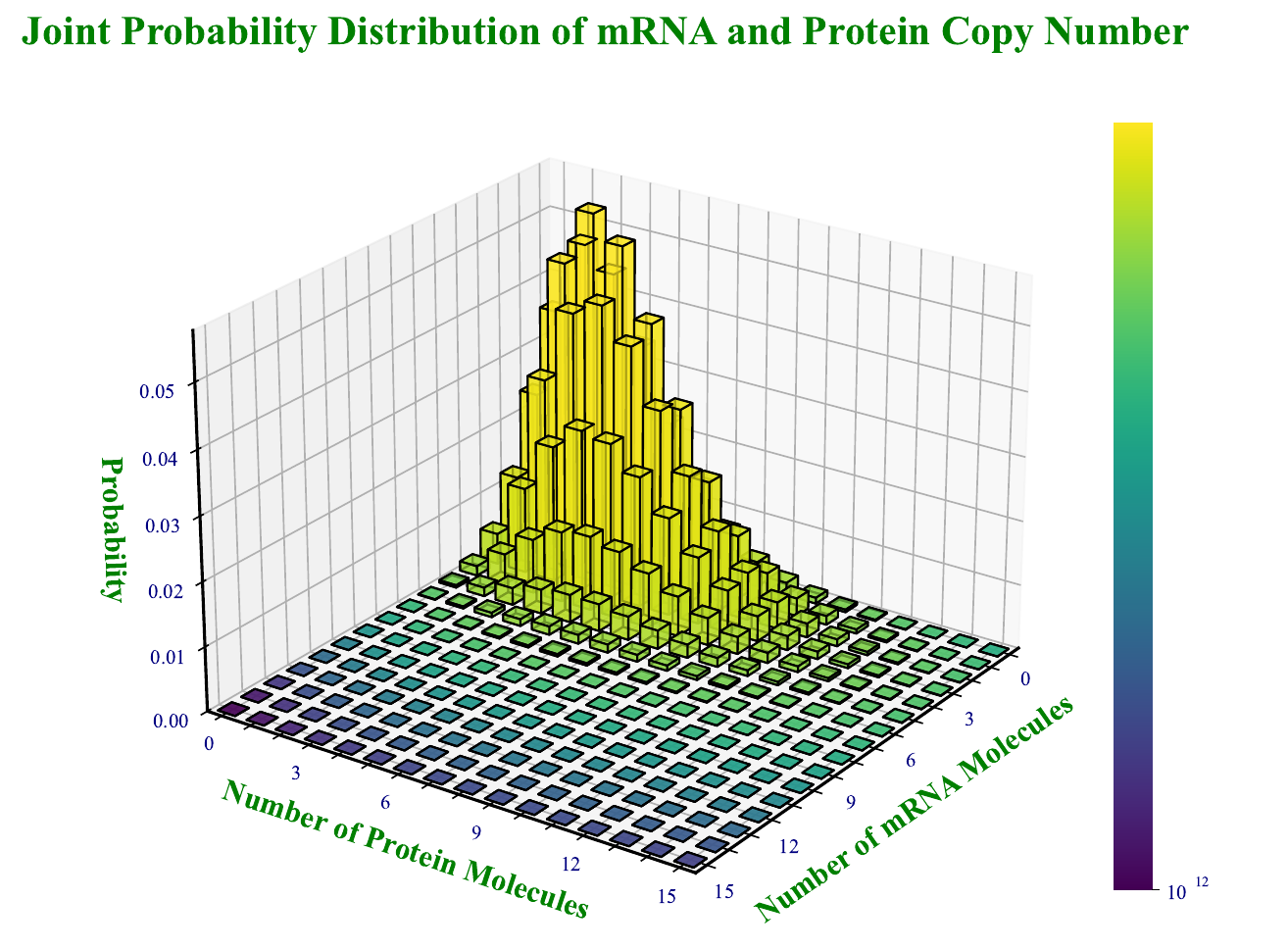}
    \includegraphics[width=1\linewidth]{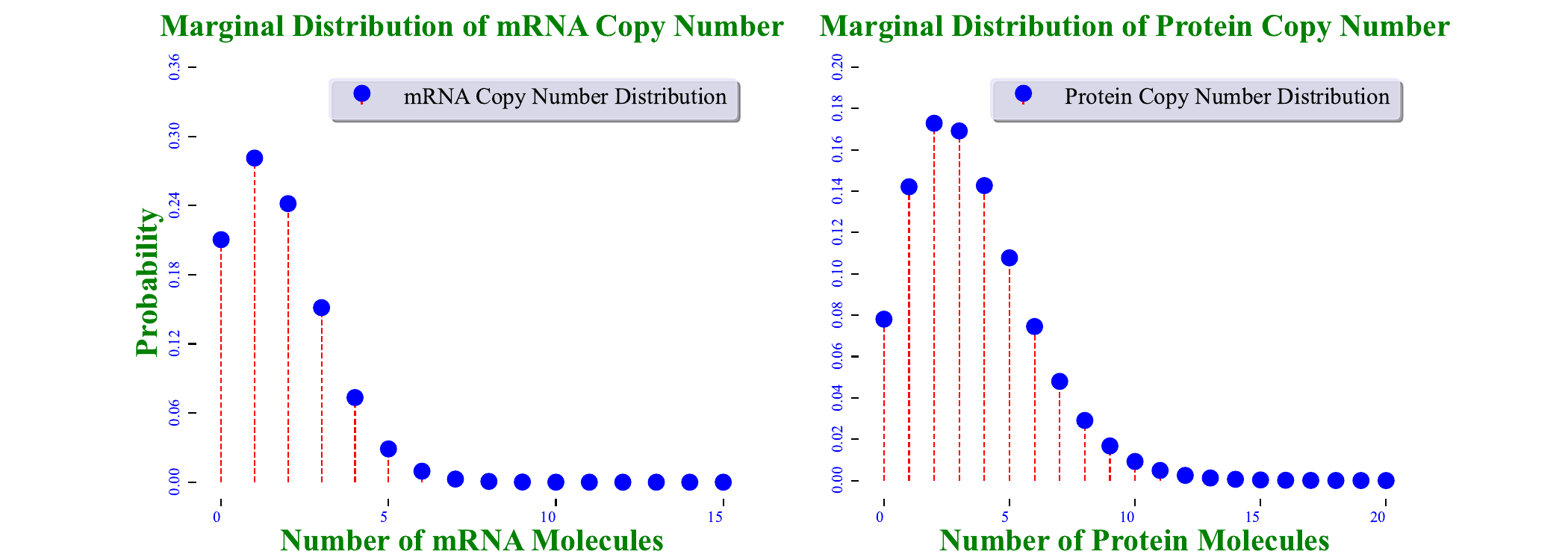}
    \caption{\textbf{Probability Distribution of mRNA and Protein Copy Number obtained through Finite State Projection Algorithm}: The parameters in the model \eqref{CME1} are the same as those in \autoref{our}. The histograms in the top panel (joint probability distribution of mRNA and protein copy number) are plotted according to FSP at $t=20$, and the truncation error is below $1\times10^{-4}$. The initial condition is $S(0)=1$, $M_1(0)=0$, and $M_2(0)=0$. The two stem plots in the bottom panel are the marginal distribution of mRNA and protein copy number, respectively. The marginal distributions are obtained directly from the joint probability distribution.}
    \label{FSP}
\end{figure}

We analyze an example using both the analytical results \eqref{BM_hierachy}, SSA, and FSP; the results are shown in \autoref{our}, \autoref{SSA}, and \autoref{FSP}, respectively. The mean squared error between the joint probability mass function in \autoref{our} and \autoref{SSA} is $1.9208\times 10^{-8}$; and between the joint probability mass function in \autoref{our} and \autoref{FSP} is $1.7685\times10^{-12}$.  See the accompanying Jupyter Notebook for details.


\subsection{Low-Order Moments}\label{sec6}
In this section, we present analytical expressions for several important low-order binomial moments, and first-order and second-order cumulants of the joint probability distribution of mRNA and protein copy number.

Using the hierarchical approach of calculating the binomial moments described in the above section, we can readily obtain explicit expressions of low-order binomial moments. For simplicity, we now stop tracking the state of the gene by considering only the coarse-grained binomial moments $B_{p,q}:=\bm{\pi}^\top\mathcal{B}_{p,q}\bm{1}$ and the probability mass function $P(m,n):=\bm{\pi}^\top\mathbb{P}(m,n)\bm{1}$ for $p,q,m,n\in\mathbb{N}$, similar to \cite{StochasticKineticsMRNA2025}. 
$\bm{1}\in\mathbb{R}^{N\times1}$ denotes the all-ones vector.
$\bm{\pi}\in\mathbb{R}^{N\times1}$ is the invariant distribution of the underlying Markov chain $(S(t))_{t\geq 0}$ characterized by $D$. Assume $D$ is irreducible. Then $\bm{\pi}$ is the unique vector satisfying $\bm{\pi}^\top D=\bm{0}_{1\times N}$ and $\bm{\pi}^\top \bm{1}=1$.
Explicit expressions of $\mathcal{B}_{p,q}$, $B_{p,q}$ with $L\leq 2$ and the detailed derivation can be found in the Supplementary Material. For comparison with binomial moments derived from models under burst approximation in \autoref{sec7}, we particularly present here the first two binomial moments of protein copy number. 
\begin{equation}\label{LowBMcoarse}
\begin{aligned}
    &B_{0,1} = \frac{v}{u\delta}\bm{\pi}^\top D_1\bm{1},\\
    &B_{0,2} = \frac{v^2}{2u\delta(u+\delta)}\bm{\pi}^\top D_1\bm{1}+\frac{v^2}{2\delta(u+\delta)}\bm{\pi}^\top D_1(u\bm{I}_N-D)^{-1}(\delta\bm{I}_N-D)^{-1}D_1\bm{1}\\&\;\;\;\;\;\;\;\;\;\;\;\;\;+\frac{v^2}{2u\delta(u+\delta)}\bm{\pi}^\top D_1(u\bm{I}_N-D)^{-1}D_1\bm{1}.
\end{aligned}
\end{equation}

Based on explicit expressions of low-order binomial moments, we can further calculate several key statistical quantities of the joint probability distribution of mRNA and protein count at steady state. Specifically, we calculate the first-order and second-order cumulants in the Supplementary Material.

\section{Accuracy of Burst Approximation}\label{sec7}
In this section, we evaluate the validity of the well-known burst approximation by comparing the analytical expressions of low-order binomial moments obtained before and after applying the burst approximation.

The burst approximation is commonly used to simplify the complete gene expression model \eqref{Reaction1}. Intuitively, the burst approximation aims to capture the experimental phenomenon that proteins are often produced in bursts. In one burst, proteins are synthesized at a relatively high rate over a short interval, followed by a long period of silence. 
One important quantitative relation underlying such burst phenomena is that, in general, mRNAs have much shorter lifetimes than proteins. In the notations of \eqref{Reaction1}, this corresponds to $u\gg\delta$. However, it has been argued that $u\gg\delta$ is far from sufficient to generate genuine translation bursts, and the bust approximation is better understood as a conceptual mathematical technique \cite{ModelsStochasticGene2005}.

Under burst approximation, mRNA molecules can be eliminated formally from the system, and only the one-dimensional probability distribution of the protein molecules needs to be determined. 
As a result, stochastic models under burst approximation are typically easier to analyze.
For models with one active gene state and multiple inactive states, a compact analytical expression for the probability mass function can be derived in terms of the generalized hypergeometric function \cite{ExactDistributionsStochastic2022,AnalyticalDistributionsStochastic2008}.
By contrast, this is generally impossible for the complete gene expression model \eqref{CME1} according to the analytical expression of the generating function (See the Supplementary
Material). 
General models with multiple gene states are also studied under burst approximation, and a recurrence relation for binomial moments can also be obtained using one-dimensional binomial moment method \cite{ExactDistributionsStochastic2022}. 
Additionally, non-Markovian models (models that are not continuous-time Markov chains) under burst approximation exist and can be studied using queueing theory \cite{IntrinsicNoiseStochastic2011}. 
We note that, according to the equivalence between queueing systems and gene expression models established in \cite{SolvingStochasticGeneexpression2024}, models under burst approximation can be interpreted as queueing systems with batch arrivals \cite{AnalysisInfiniteserverQueue2002}. This perspective may explain the significant convenience introduced by the burst approximation.

In general, stochastic gene expression models under burst approximation possess independent theoretical value. However, in this article, we focus on the validity of burst approximation when replacing the original reaction system \eqref{Reaction1} with an alternative. Therefore, the general setting is not introduced here. 
A systematic analysis of general stochastic gene expression models under burst approximation will be reported in a forthcoming paper \cite{StochasticKineticsProtein2025}. 

Here, we approximate \eqref{Reaction1} with the following reaction system \eqref{Reaction2}.
\begin{equation}\label{Reaction2}
\begin{aligned}
    &\ce{\mathcal{S}_i ->[$a_{i,j}$] \mathcal{S}_j}\;\;(i\neq j,\;\;1\leq i,j\leq N)\\
    &\ce{\mathcal{S}_i ->[$b_{i,j}^{[r]}$] \mathcal{S}_j + r \cdot \textbf{Protein}}\;\;(1\leq i,j\leq N,\;\;r=1,2,\cdots)\\
    &\ce{\textbf{Protein} ->[$\delta$] $\emptyset$}\\
\end{aligned}
\end{equation}
In \eqref{Reaction2}, the species $\mathcal{S}_i\;(1\leq i\leq N)$, \textbf{Protein}, and the parameters $a_{i,j}\;(i\neq j,\;\;1\leq i,j\leq N)$, $\delta$ carry the same meaning as in \eqref{Reaction1}. Additionally, under the notations in \eqref{Reaction1}, $b_{i,j}^{[r]}\;(1\leq i,j\leq N,\;\;r=1,2,\cdots)$ are given by the following relation.
\begin{equation}\label{appr}
\begin{aligned}
    b_{i,j}^{[r]}:=\left(\frac{v}{u+v}\right)^r\left(1-\frac{v}{u+v}\right)b_{i,j},\;\;1\leq i,j\leq N,\;\;r=1,2,\cdots.
\end{aligned}
\end{equation}

The relation \eqref{appr} can be interpreted as follows. 
Once transcribed, an mRNA molecule is subject to two competing reaction pathways, namely, hydrolysis and translation. More specifically, this can be seen as the competing binding of decay complexes that promote hydrolysis, and recruitment of initiation factors that engage the ribosome for translation. Since the probability of initiating translation rather than hydrolysis is $v/(u+v)$, the number of protein molecules produced from a single mRNA molecule follows a geometric distribution with parameter $u/(u+v)$. Hence, \eqref{appr} readily follows. We note that the geometrically distributed burst size is consistent with experimental observations \cite{StochasticProteinExpression2006}.

With the notation introduced above, the first-order and second-order binomial moments of protein copy number in \eqref{Reaction2} are
\begin{equation}\label{LowBMburst}
\begin{aligned}
    \widetilde{B_1}&=\frac{v}{u\delta}\bm{\pi}^\top D_1\bm{1},\\
    \widetilde{B_2}&=\frac{v^2}{2u^2\delta}\bm{\pi}^\top D_1\bm{1}+\frac{v^2}{2u^2\delta}\bm{\pi}^\top D_1\left(\delta\bm{I}_N-D\right)^{-1}D_1\bm{1}.
\end{aligned}
\end{equation}
The detailed derivation can be found in \cite{StochasticKineticsProtein2025}.

\begin{figure}[t!]
    \centering
    \includegraphics[width=1\linewidth]{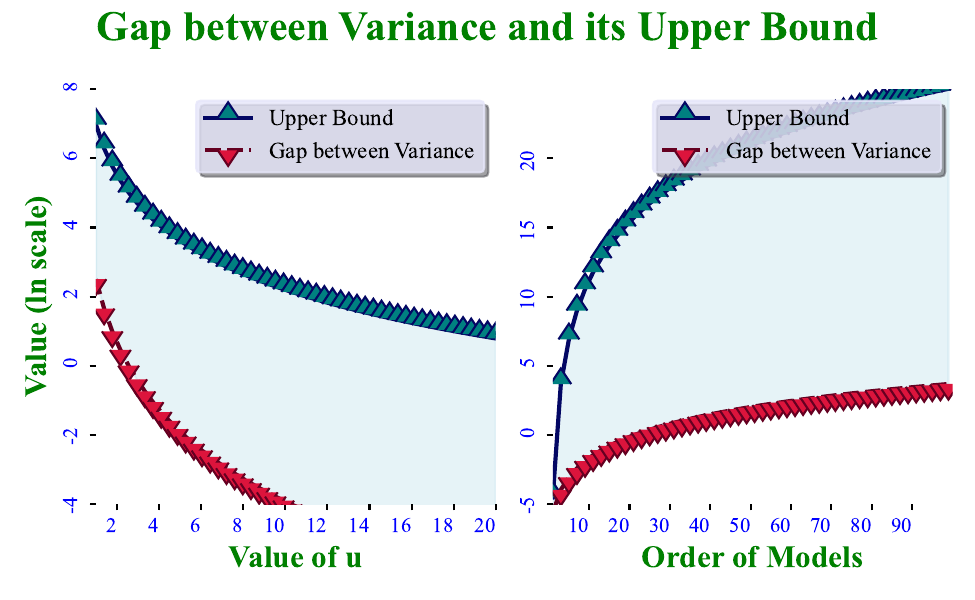}
    \caption{\textbf{Gap between Variance and its Upper Bound}: In the left panel, all the parameters except for $u$ are the same as those in \autoref{our} and are fixed. While $u$ varies in $[1,20]$, we compute the discrepancy between the variances of the protein copy number according to \eqref{LowBMcoarse} and \eqref{LowBMburst}, and also evaluate its upper bound given by \eqref{accuracy}. In the right panel, $u=10$, $v=2$, and $\delta=1$ are fixed, while $D_0$ and $D_1$ take the form $a_{i,j}=0\;(i\neq j)$ and $b_{i,j}=i$. For the order of the model $1\leq N\leq 100$, we compute the discrepancy between the variances of the protein copy number according to \eqref{LowBMcoarse} and \eqref{LowBMburst}, and also evaluate its upper bound given by \eqref{accuracy}. Note that both illustrations are plotted in $\ln(\cdot)$ scale, and that the discrepancy between variances equals two times the discrepancy between second-order binomial moments.}
    \label{bound}
\end{figure}

Compared with \eqref{LowBMcoarse}, it is worth noting that the first-order binomial moment, equivalently the expectation of the protein copy number, remains exact under the burst approximation. By contrast, the second-order binomial moment is explicitly altered when burst approximation is applied. In particular, an upper bound can be derived for the difference between the binomial moments obtained from the two models, using techniques from functional analysis and an analogue to Theorem 4.1.2 in \cite{MatrixComputations2013}. Specifically, we have
\begin{equation}\label{accuracy}
\begin{aligned}
    \lvert B_{0,2}- \widetilde{B_2}\rvert\leq&\frac{v^2}{2u^2(u+\delta)}\lVert D_1\rVert_{\infty}
   +\frac{v^2}{u^2\delta(u+\delta)}\lVert D_1\rVert_{\infty}^2
   +\frac{v^2}{2u^2\delta^2}\lVert D_1\rVert_{\infty}^2\lVert D\rVert_{\infty}.
\end{aligned}
\end{equation}
Note that the infinity norm of a matrix, denoted by $\lVert \cdot\rVert_\infty$, is the maximum absolute row sum of this matrix. The proof can be found in the Supplementary Material. According to \eqref{accuracy}, the difference converges to zero at the rate of $O(u^{-2})$ as $u\rightarrow\infty$ while the other parameters are fixed. In general, $u/\delta\gg1$ does not guarantee the validity of the burst approximation. In the left panel of \autoref{bound}, both the upper bound given by \eqref{accuracy} and the gap between the variances converge to zero as $u$ grows and all other parameters remain the same. However, according to the right panel of \autoref{bound}, although $u/\delta\gg1$, the gap between variances of the protein copy number in complete gene expression models and the corresponding surrogate models can be arbitrarily large, indicating that the burst approximation can be inaccurate in some cases.

\section{Discussion and Conclusion}
In this paper, we establish an effective approach to analyzing stochastic gene expression models without resorting to burst approximation. 
Analytical expressions of binomial moments of mRNA and protein counts can be obtained up to any order at steady state in a hierarchical manner. 
Numerical computation of binomial moments is both fast and accurate.
Subsequently, the joint probability mass function can be reconstructed. 
In particular, based on analytical expressions of low-order cumulants, we can rigorously evaluate the validity of the burst approximation for general gene expression models. 

We first note that the asymptotic behavior of binomial moments and probability mass function needs further analysis. In \cite{StochasticKineticsMRNA2025}, we derive elegant upper bounds for binomial moments and probability mass function of mRNA copy number, which are $B_{p,0},\;p\in\mathbb{N}$ and $\sum_{n=0}^\infty P(m,n),\;m\in\mathbb{N}$ in this article, respectively. 
However, deriving concise upper bounds for $B_{p,q},\;p,q\in\mathbb{N}$ or $P(m,n),\;m,n\in\mathbb{N}$ appears substantially more involved. 
Note that asymptotic analysis is crucial to proving the convergence of \eqref{reconstruct} and designing an appropriate truncation strategy. A rigorous treatment is left to future work.

Finally, we examine the gene-state topology from a multiscale modeling perspective. 
Given the flexibility of our general model \eqref{Reaction1}, it becomes essential to predefine the number of states $N$ and the transition structure, namely $D_0$ and $D_1$.
Otherwise, the model may yield highly pathological results and nonphysical predictions. 
Fortunately, from a multiscale modeling perspective, this task may be accomplished by molecular dynamics (MD) simulations \cite{StatisticalMechanicsTheory2023}. 
At the atomistic level, dynamics of macromolecules, such as DNA and proteins, can be accurately captured using MD simulations. Physically, each gene state corresponds to a metastable configuration of the underlying macromolecule, namely DNA in this problem. 
Thus, the Markov chain of gene state is a coarse-grained representation of the underlying dynamics. Designing the gene-state topology essentially amounts to constructing a Markov state model (MSM) from MD trajectories, which is a standard method for analyzing MD data and understanding conformational transitions \cite{IdentificationSlowMolecular2013,MetastabilityMarkovState2013}.
Constructing a MSM typically involves two steps. The first step is to project the high-dimensional MD data, namely the coordinates of all atoms in a macromolecule, onto a relatively lower-dimensional latent space. The above dimensionality reduction relies on specifying collective variables (CVs), which map the complete configurations to low-dimensional representations but preserve non-trivial dynamical behaviors of the molecule. 
Next, one clusters the data in CV space and uses cluster centroids as discrete states of a Markov chain, whose transition rates are estimated at a given lag time. Nevertheless, constructing CVs remains a challenge for real-world macromolecules, although machine learning techniques have been incorporated to develop many useful methods \cite{CollectiveVariableDiscovery2022,UnifiedFrameworkMachine2023}. Therefore, further work is needed to develop a physically grounded, bottom-up stochastic gene expression model.

\section*{Acknowledgments}
Y.L. thanks Pinchen Xie and Ruiqi Gao for fruitful discussions. Y.L. was supported by Natural Science Foundation of China (125B10002), FDUROP (Fudan Undergraduate Research Opportunities Program) (24260), and Shanghai Undergraduate Training Program on Innovation and Entrepreneurship grant (S202510246532). Y.Z. was supported by National Key R$\&$D Program of China (2024YFA1012401), the Science and Technology Commission of Shanghai Municipality (23JC1400501), and Natural Science Foundation of China (12241103).
\section*{Conflict of Interest}
The authors have no conflicts to disclose.
\section*{Code Availability}
The Python code will be released in a public repository soon.


\appendix

\section{Matrix-form Chemical Master Equation and Generating Function Method}
In this section, we show that the chemical master equation \eqref{CME1} can be equivalently reformulated as the partial differential equation system \eqref{partial1} using the standard generating function method. 

Recall the definition of $a_{i,i}$, the chemical master equation of \eqref{Reaction1}, namely, \eqref{CME1}, can be rearranged as
\begin{equation}\label{CME2}
\begin{aligned}
\frac{\partial}{\partial t}\mathbb{P}_{i,j}(m,n;t)&=\sum_{s=1}^Na_{s,j}\mathbb{P}_{i,s}(m,n;t)+\sum_{s=1}^Nb_{s,j}\mathbb{P}_{i,s}(m-1,n;t)\\
&+mv\mathbb{P}_{i,j}(m,n-1;t)+(m+1)u\mathbb{P}_{i,j}(m+1,n;t)\\&+(n+1)\delta\mathbb{P}_{i,j}(m,n+1;t)-(mu+n\delta+mv)\mathbb{P}_{i,j}(m,n;t).
\end{aligned}
\end{equation}
Given $m$, $n$ and $t$, assemble the entries $\mathbb{P}_{i,j}(m,n;t)$ into a matrix according to the subindex $(i,j)$, obtaining the matrix-form probability mass function $\mathbb{P}(m,n;t)\in\mathbb{R}^{N\times N}$. With this notation, \eqref{CME2} takes the compact form
\begin{equation}\label{CME_Matrix}
\begin{aligned}
\frac{\partial}{\partial t}\mathbb{P}(m,n;t)&=\mathbb{P}(m,n;t)D_0+\mathbb{P}(m-1,n;t)D_1\\&+mv\mathbb{P}(m,n-1;t)+(m+1)u\mathbb{P}(m+1,n;t)\\&+(n+1)\delta\mathbb{P}(m,n+1;t)-(mu+n\delta+mv)\mathbb{P}(m,n;t).
\end{aligned}
\end{equation}

Then we refer to standard generating function method. Recall that the matrix-form generating function $\mathcal{G}(z,w;t)$ is defined in the main text by \eqref{generating1}. The chemical master equation \eqref{CME_Matrix} can be equivalently converted into the partial differential equation system.

Converting \eqref{CME2} into matrix-form and using generating function, the original chemical master equation system can be neatly expressed as the following partial differential equation system \eqref{partial1}.

\section{Differential Equations for the Binomial Moments}
In this section, we establish the relations between binomial moments and the probability mass function, and derive the hierarchy of ordinary differential equations governing binomial moments.

According to the formal definition \eqref{BM} of binomial moments $\mathcal{B}_{p,q}(t)$, we can now verify the equality \eqref{related}, stating that binomial moments are Taylor coefficients of the generating function $\mathcal{G}(z,w;t)$ expanded around $z=1,w=1$. Note that, since $z=s+1$ and $w=r+1$, 
\begin{equation}\label{related_detail}
\begin{aligned}
    \mathcal{G}(z,w;t)&\equiv\sum_{m=0}^\infty\sum_{n=0}^\infty z^mw^n\mathbb{P}(m,n;t)=\sum_{m=0}^\infty\sum_{n=0}^\infty (s+1)^m(r+1)^n\mathbb{P}(m,n;t)\\&
    =\sum_{m=0}^\infty\sum_{p=0}^m\sum_{n=0}^\infty\sum_{q=0}^n\binom{m}{p}\binom{n}{q}s^pr^q\mathbb{P}(m,n;t)\\&
    =\sum_{p=0}^\infty\sum_{m=p}^\infty\sum_{q=0}^\infty\sum_{n=q}^\infty\binom{m}{p}\binom{n}{q}s^pr^q\mathbb{P}(m,n;t)\\&
    =\sum_{p=0}^\infty\sum_{q=0}^\infty s^pr^q\left[\sum_{m=p}^\infty\sum_{n=q}^\infty\binom{m}{p}\binom{n}{q}\mathbb{P}(m,n;t)\right]\\&
    \equiv\sum_{p=0}^\infty\sum_{q=0}^\infty s^pr^q\mathcal{B}_{p,q}(t).
\end{aligned}
\end{equation}

Likewise, by inverting the above derivation, we arrive at the relation \eqref{reconstruct} showing how binomial moments determine the probability mass function. 
\begin{equation}
\begin{aligned}
\sum_{p=0}^\infty\sum_{q=0}^\infty s^pr^q\mathcal{B}_{p,q}(t)&\equiv\sum_{p=0}^\infty\sum_{q=0}^\infty (z-1)^p(w-1)^q\mathcal{B}_{p,q}(t)\\&=\sum_{p=0}^\infty\sum_{m=0}^p\sum_{q=0}^\infty\sum_{n=0}^q(-1)^{p+q-m-n}\binom{p}{m}\binom{q}{n}z^mw^n\mathcal{B}_{p,q}(t)\\&
=\sum_{m=0}^\infty\sum_{p=m}^\infty\sum_{n=0}^\infty\sum_{q=n}^\infty(-1)^{p+q-m-n}\binom{p}{m}\binom{q}{n}z^mw^n\mathcal{B}_{p,q}(t)\\&
=\sum_{m=0}^\infty\sum_{n=0}^\infty z^mw^n\left[\sum_{p=m}^\infty\sum_{q=n}^\infty(-1)^{p+q-m-n}\binom{p}{m}\binom{q}{n}\mathcal{B}_{p,q}(t)\right]\\&
=\sum_{m=0}^\infty\sum_{n=0}^\infty z^mw^n\left[\sum_{p=m}^\infty\sum_{q=n}^\infty(-1)^{p+q+m+n}\binom{p}{m}\binom{q}{n}\mathcal{B}_{p,q}(t)\right].
\end{aligned}
\end{equation}

By the identity given in \eqref{related}, the above expression can equivalently be expressed as $\mathcal{G}(z,w;t)\equiv\sum_{m=0}^\infty\sum_{n=0}^\infty z^mw^n\mathbb{P}(m,n;t)$. From the uniqueness of the Taylor coefficients, it follows that
\begin{equation}
\begin{aligned}
\mathbb{P}(m,n;t)=\sum_{p=m}^\infty\sum_{q=n}^\infty(-1)^{p+q+m+n}\binom{p}{m}\binom{q}{n}\mathcal{B}_{p,q}(t).
\end{aligned}
\end{equation}

To obtain the differential equation system governing binomial moments, we now substitute \eqref{related} into the partial differential equation system \eqref{partial1}, and it follows that 
\begin{equation}\label{BM_1}
\begin{aligned}
\sum_{p=0}^\infty\sum_{q=0}^\infty s^pr^q&\frac{\mathrm{d}}{\mathrm{d} t}\mathcal{B}_{p,q}(t)=\sum_{p=0}^\infty\sum_{q=0}^\infty s^pr^q\mathcal{B}_{p,q}(t)D_0+(s+1)\sum_{p=0}^\infty\sum_{q=0}^\infty s^pr^q\mathcal{B}_{p,q}(t)D_1\\&+\left[-us+v(s+1)r\right]p\sum_{p=1}^\infty\sum_{q=0}^\infty s^{p-1}r^q\mathcal{B}_{p,q}(t)-\delta rq\sum_{p=0}^\infty\sum_{q=1}^\infty s^{p}r^{q-1}\mathcal{B}_{p,q}(t).
\end{aligned}
\end{equation}

Arranging the right hand side of \eqref{BM_1} according to the order of $s$ and $r$, we obtain a hierarchy of ordinary differential equations governing binomial moments, namely, 
\begin{equation}\label{ODE}
\begin{aligned}
\frac{\mathrm{d}}{\mathrm{d} t}\mathcal{B}_{p,q}(t)&=\mathcal{B}_{p,q}(t)(D_0+D_1-up\bm{I}_N-\delta q\bm{I}_N)+\mathcal{B}_{p-1,q}(t)D_1\\&+vp\mathcal{B}_{p,q-1}(t)+v(p+1)\mathcal{B}_{p+1,q-1}(t),\;\;p,q\in\mathbb{N}.
\end{aligned}
\end{equation}

Consider only the steady state, we may the time-derivatives in \eqref{ODE} to zero and thus obtain \eqref{BM_hierachy}. Recall that $D_0+D_1=D$.


\section{Low-Order Binomial Moments}
Based on the recurrence relation \eqref{BM_hierachy}, analytical expressions for low-order binomial moments can be readily obtained, for example, 
\begin{equation}\label{LowBM}
\begin{aligned}
    &\mathcal{B}_{1,0} = D_1(u\bm{I}_N-D)^{-1}\\
    &\mathcal{B}_{0,1} = vD_1(u\bm{I}_N-D)^{-1}(\delta\bm{I}_N-D)^{-1}\\
    &\mathcal{B}_{2,0}=D_1(u\bm{I}_N-D)^{-1}D_1(2u\bm{I}_N-D)^{-1}\\
    &\mathcal{B}_{1,1}=vD_1(u\bm{I}_N-D)^{-1}\left[I_N+(\delta\bm{I}_N-D)^{-1}D_1+2D_1(2u\bm{I}_N-D)^{-1}\right](u\bm{I}_N+\delta\bm{I}_N-D)^{-1}\\
    &\mathcal{B}_{0,2}=v^2D_1(u\bm{I}_N-D)^{-1}\left[I_N+(\delta\bm{I}_N-D)^{-1}D_1+2D_1(2u\bm{I}_N-D)^{-1}\right]\\&\;\;\;\;\;\;\;\;\;\;\;\;\;\;\times(u\bm{I}_N+\delta\bm{I}_N-D)^{-1}(2\delta\bm{I}_N-D)^{-1}.
\end{aligned}
\end{equation}

To obtain the expressions in \eqref{LowBMcoarse}, recall that the $D$ is a $Q$-matrix, satisfying $D\bm{1}=\bm{0}_{N\times1}$. Therefore, 
\begin{equation}\label{trick}
    \left(k\bm{I}_N-D\right)^{-1}\bm{1}=\frac{1}{k}\bm{1},\;\;k>0.
\end{equation}

Using \eqref{trick}, it follows readily from \eqref{LowBM} that 
\begin{equation}
\begin{aligned}
    &B_{1,0} = \frac{1}{u}\bm{\pi}^\top D_1\bm{1},\\&B_{0,1} = \frac{v}{u\delta}\bm{\pi}^\top D_1\bm{1},\\&B_{2,0} = \frac{1}{2u}\bm{\pi}^\top D_1(u\bm{I}_N-D)^{-1}D_1\bm{1},\\
    &B_{1,1} = \frac{v}{u(u+\delta)}\bm{\pi}^\top D_1\bm{1}+\frac{v}{u+\delta}\bm{\pi}^\top D_1(u\bm{I}_N-D)^{-1}(\delta\bm{I}_N-D)^{-1}D_1\bm{1}\\&\;\;\;\;\;\;\;\;\;\;\;\;\;+\frac{v}{u(u+\delta)}\bm{\pi}^\top D_1(u\bm{I}_N-D)^{-1}D_1\bm{1},\\
    &B_{0,2} = \frac{v^2}{2u\delta(u+\delta)}\bm{\pi}^\top D_1\bm{1}+\frac{v^2}{2\delta(u+\delta)}\bm{\pi}^\top D_1(u\bm{I}_N-D)^{-1}(\delta\bm{I}_N-D)^{-1}D_1\bm{1}\\&\;\;\;\;\;\;\;\;\;\;\;\;\;+\frac{v^2}{2u\delta(u+\delta)}\bm{\pi}^\top D_1(u\bm{I}_N-D)^{-1}D_1\bm{1}.
\end{aligned}
\end{equation}

\section{Proof of \eqref{accuracy}}
\begin{equation}
\begin{aligned}
   \lvert B_{0,2}- \widetilde{B_2}\rvert&=\lvert \frac{v^2}{2u\delta(u+\delta)}\bm{\pi}^\top D_1\bm{1}+\frac{v^2}{2\delta(u+\delta)}\bm{\pi}^\top D_1(u\bm{I}_N-D)^{-1}(\delta\bm{I}_N-D)^{-1}D_1\bm{1}\\&
   +\frac{v^2}{2u\delta(u+\delta)}\bm{\pi}^\top D_1(u\bm{I}_N-D)^{-1}D_1\bm{1}- \frac{v^2}{2u^2\delta}\bm{\pi}^\top D_1\bm{1}-\frac{v^2}{2u^2\delta}\bm{\pi}^\top D_1\left(\delta\bm{I}_N-D\right)^{-1}D_1\bm{1}\rvert\\&
   \leq\lvert\frac{v^2}{2u\delta(u+\delta)}\bm{\pi}^\top D_1\bm{1}-\frac{v^2}{2u^2\delta}\bm{\pi}^\top D_1\bm{1}\rvert\\&
   +\lvert\frac{v^2}{2\delta(u+\delta)}\bm{\pi}^\top D_1(u\bm{I}_N-D)^{-1}(\delta\bm{I}_N-D)^{-1}D_1\bm{1}-\frac{v^2}{2u^2\delta}\bm{\pi}^\top D_1\left(\delta\bm{I}_N-D\right)^{-1}D_1\bm{1}\rvert\\&
   +\frac{v^2}{2u\delta(u+\delta)}\lvert\bm{\pi}^\top D_1(u\bm{I}_N-D)^{-1}D_1\bm{1}\rvert\\&
   \leq\frac{v^2}{2u^2(u+\delta)}\lvert\bm{\pi}^\top D_1\bm{1}\rvert\\&
   +\frac{v^2}{2u(u+\delta)}\lvert\bm{\pi}^\top D_1(u\bm{I}_N-D)^{-1}(\delta\bm{I}_N-D)^{-1}D_1\bm{1}\rvert\\&
   +\frac{v^2}{2u\delta}\lvert\bm{\pi}^\top D_1(u\bm{I}_N-D)^{-1}(\delta\bm{I}_N-D)^{-1}D_1\bm{1}-\bm{\pi}^\top D_1\left(\frac{1}{u}\bm{I}_N\right)\left(\delta\bm{I}_N-D\right)^{-1}D_1\bm{1}\rvert\\&
   +\frac{v^2}{2u\delta(u+\delta)}\lvert\bm{\pi}^\top D_1(u\bm{I}_N-D)^{-1}D_1\bm{1}\rvert\\&
   \leq\frac{v^2}{2u^2(u+\delta)}\lVert \bm{\pi}\rVert_{1}\lVert D_1\rVert_{\infty}\lVert\bm{1}\rVert_{\infty}\\&
   +\frac{v^2}{2u(u+\delta)}\lVert \bm{\pi}\rVert_{1}\lVert D_1\rVert_{\infty}\lVert (u\bm{I}_N-D)^{-1}\rVert_{\infty}\lVert (\delta\bm{I}_N-D)^{-1}\rVert_{\infty}\lVert D_1\rVert_{\infty}\lVert\bm{1}\rVert_{\infty}\\&
   +\frac{v^2}{2u\delta}
   \lVert \bm{\pi}\rVert_{1}\lVert D_1\rVert_{\infty}\lVert (u\bm{I}_N-D)^{-1}-\frac{1}{u}\bm{I}_N\rVert_{\infty}\lVert (\delta\bm{I}_N-D)^{-1}\rVert_{\infty}\lVert D_1\rVert_{\infty}\lVert\bm{1}\rVert_{\infty}\\&
   +\frac{v^2}{2u\delta(u+\delta)}\lVert\bm{\pi}\rVert_{1}\lVert D_1\rVert_{\infty}\lVert (u\bm{I}_N-D)^{-1}\rVert_{\infty}\lVert D_1\rVert_{\infty}\lVert\bm{1}\rVert_{\infty}\\&
   \leq\frac{v^2}{2u^2(u+\delta)}\lVert D_1\rVert_{\infty}
   +\frac{v^2}{2u^2\delta(u+\delta)}\lVert D_1\rVert_{\infty}^2\\&
   +\frac{v^2}{2u^2\delta^2}\lVert D_1\rVert_{\infty}^2\lVert D\rVert_{\infty}+\frac{v^2}{2u^2\delta(u+\delta)}\lVert D_1\rVert_{\infty}^2\\&
   =\frac{v^2}{2u^2(u+\delta)}\lVert D_1\rVert_{\infty}
   +\frac{v^2}{u^2\delta(u+\delta)}\lVert D_1\rVert_{\infty}^2+\frac{v^2}{2u^2\delta^2}\lVert D_1\rVert_{\infty}^2\lVert D\rVert_{\infty}. 
\end{aligned}
\end{equation}

In the penultimate inequality, we applied H\"older's inequality. In the last inequality, we used an analogue to Theorem 4.1.2 in \cite{MatrixComputations2013}, stating that 
\begin{theorem}\label{fromMC}
Let $C=(c_{i,j})_{N\times N}$ be a $N\times N$ strictly row diagonally dominant matrix with $\Theta:=\min_{1\leq i\leq N}\left(\mid c_{i,i}\mid-\sum_{j\neq i}\mid c_{i,j}\mid\right)$. Then $\lVert C^{-1}\rVert_\infty\leq \Theta^{-1}$. 
\end{theorem}
Since $D$ is the generator of a continuous-time Markov chain, for any $\lambda>0$, the matrix $\lambda\bm{I}_N-D$ is strictly row diagonally dominant, with each row summing to $\lambda$. Therefore $\lVert(\lambda\bm{I}_N-D)^{-1}\rVert_\infty\leq 1/\lambda$.

Note also that 
\begin{equation}
\begin{aligned}
    \lVert (u\bm{I}_N-D)^{-1}-\frac{1}{u}\bm{I}_N\rVert_{\infty}&=\lVert (u\bm{I}_N-D)^{-1}\left[\bm{I}_N-\frac{1}{u}(u\bm{I}_N-D)\right]\rVert_{\infty}\\&
    \leq\lVert (u\bm{I}_N-D)^{-1}\rVert_{\infty}\lVert D\rVert_{\infty}\leq\frac{1}{u}\lVert D\rVert_{\infty}.
\end{aligned}
\end{equation}

\section{Analytical Expression of Generating Function}
In this section, we verify that \eqref{solution1} is the unique solution to \eqref{partial1} under initial condition $\mathcal{G}(z,w;0)=\bm{I}_N$, therefore presenting the explicit expression of generating function. 

We claim that the unique solution to \eqref{partial1} under initial condition $\mathcal{G}(z,w;0)=\bm{I}_N$ is as follow.
\begin{equation}\label{solution1}
\begin{aligned}
\mathcal{G}(z,w;t)=\bm{I}_N+\int_0^tH(t_1)\mathrm{d}t_1+\sum_{k=2}^\infty\int_0^t\int_0^{t_1}\cdots\int_0^{t_{k-1}}H(t_k)H(t_{k-1})\cdots H(t_1) \mathrm{d}t_{k}\cdots\mathrm{d}t_2\mathrm{d}t_1,
\end{aligned}
\end{equation}
where $H(s;z,w;t):=D_0+h(s;z,w;t)D_1\in\mathbb{R}^{N\times N}$ and $h(s;z,w;t)$ is a real-valued function whose expression is given below.
\begin{equation}\label{solutionh}
\begin{aligned}
h(s;z,w;t) &:= 
  \exp\!\left[ us + \tfrac{v}{\delta}(1-w)\,
             \exp\!\bigl(\delta s-\delta t) \right] \\
&\quad \times \Biggl\{
      z \exp\!\left[-ut - \tfrac{v}{\delta}(1-w)\right] \\
&\qquad + \frac{u}{\delta}
         \left[ \frac{v}{\delta}(1-w) \right]^{u/\delta}
         \exp(-ut) \\
&\qquad\quad \times \Bigl[
             \Gamma\!\left(-\tfrac{u}{\delta};
             \tfrac{v}{\delta}(1-w)\exp(\delta s-\delta t)\right)
             - \Gamma\!\left(-\tfrac{u}{\delta};
             \tfrac{v}{\delta}(1-w)\right)
           \Bigr]
    \Biggr\}.
\end{aligned}
\end{equation}
In \eqref{solutionh}, $\Gamma(\gamma;x):=\int_x^\infty y^{\gamma-1}\mathrm{e}^{-y}\mathrm{d}y$ is the incomplete gamma function.

It is obvious that \eqref{solution1} satisfies the initial condition $\mathcal{G}(z,w;0)=\bm{I}_N$. Therefore we only need to verify that $\mathcal{G}(z,w;t)$ given by \eqref{solution1} satisfies the partial differential equation \eqref{partial1}. The verification is straightforward but tedious, and we proceed in three steps.
\subsection{Abbreviations}
For simplicity, we first introduce some abbreviations for the calculation.
\begin{equation}
\begin{aligned}
    &a(z,w):=u(1-z)+vz(w-1),\; b(w):=\delta(1-w),\\&
    \theta:= \text{any one of }\{z,w,t\},\\&c(w):=\frac{v}{\delta}(w-1),\; \alpha:=-\frac{u}{\delta},\; X(s;w;t):=-c(w)e^{\delta(s-t)},\\&
    A(s;w;t):=\exp\big[us+X(s;w;t)\big],\;E(w;t):=\exp\big[-ut+c(w)\big],\\&M(w;t):=\frac{u}{\delta}\mathrm{e}^{-ut}[-c(w)]^{u/\delta},\;
    H(s;w;t):=\Gamma(\alpha;X(s;w;t))-\Gamma(\alpha,-c(w)).
\end{aligned}
\end{equation}
Note that with the above abbreviations, we have from \eqref{solutionh}
\begin{equation}\label{abbr}
\begin{aligned}
    h(s;z,w;t)=A(s;w;t)\left[E(w;t)\,z+M(w;t)H(s;w;t)\right],\;\;0\leq s\leq t.
\end{aligned}
\end{equation}

\subsection{Derivatives of Peano-Baker Series}
We now take derivatives with respect to $t$ in \eqref{solution1}, which takes the form of a Peano-Baker Series or time-ordering operator. 

Note that
\begin{equation}
\begin{aligned}
    \frac{\partial}{\partial t}\int_0^tH(t_1)\mathrm{d}t_1=H(t)+\int_0^t\frac{\partial}{\partial t}H(t_1)\mathrm{d}t_1=D_0+zD_1+\int_0^t\frac{\partial}{\partial t}H(t_1)\mathrm{d}t_1.
\end{aligned}
\end{equation}
Similarly, for $k\geq 2$,
\begin{equation}
\begin{aligned}
    &\frac{\partial}{\partial t}\int_0^t\int_0^{t_1}\cdots\int_0^{t_{k-1}}H(t_k)H(t_{k-1})\cdots H(t_1) \mathrm{d}t_{k}\cdots\mathrm{d}t_2\mathrm{d}t_1\\&=\int_0^{t}\int_0^{t_2}\cdots\int_0^{t_{k-1}}H(t_k)H(t_{k-1})\cdots H(t_2)H(t) \mathrm{d}t_{k}\cdots\mathrm{d}t_2\\&+\int_0^t\frac{\partial}{\partial t}\left[\int_0^{t_1}\cdots\int_0^{t_{k-1}}H(t_k)H(t_{k-1})\cdots H(t_1) \mathrm{d}t_{k}\cdots\mathrm{d}t_2\right]\mathrm{d}t_1\\&
    =\int_0^{t}\int_0^{t_2}\cdots\int_0^{t_{k-1}}H(t_k)H(t_{k-1})\cdots H(t_2) \mathrm{d}t_{k}\cdots\mathrm{d}t_2(D_0+zD_1)\\&
    +\int_0^t\int_0^{t_1}\cdots\int_0^{t_{k-1}}\frac{\partial}{\partial t}\left[H(t_k)H(t_{k-1})\cdots H(t_1)\right] \mathrm{d}t_{k}\cdots\mathrm{d}t_2\mathrm{d}t_1.
\end{aligned}
\end{equation}
In addition, for $k\geq 2$,
\begin{equation}\label{chain}
\begin{aligned}
    \frac{\partial}{\partial\theta}\left[H(t_k)H(t_{k-1})\cdots H(t_1)\right]=&\left[\frac{\partial}{\partial\theta}H(t_k)\right]H(t_{k-1})\cdots H(t_1)+H(t_k)\left[\frac{\partial}{\partial\theta}H(t_{k-1})\right]\cdots H(t_1)\\&+\cdots+H(t_k)H(t_{k-1})\cdots \left[\frac{\partial}{\partial\theta}H(t_1)\right].
\end{aligned}
\end{equation}
Therefore,
\begin{equation}
\begin{aligned}
    \frac{\partial}{\partial t}\mathcal{G}(z,w;t)&=\mathcal{G}(z,w;t)(D_0+zD_1)\\&
    +\int_0^t\frac{\partial}{\partial t}H(t_1)\mathrm{d}t_1+\sum_{k=2}^\infty\int_0^t\int_0^{t_1}\cdots\int_0^{t_{k-1}}\frac{\partial}{\partial t}\left[H(t_k)H(t_{k-1})\cdots H(t_1)\right] \mathrm{d}t_{k}\cdots\mathrm{d}t_2\mathrm{d}t_1.
\end{aligned}
\end{equation}

Using \eqref{chain}, the verification of the differential equation system \eqref{partial1} can be reduced to proving that the real-valued multi-variable function $h(s;z,w;t)$ satisfies the following differential equation. 
\begin{equation}\label{partical3}
\begin{aligned}
    \partial_t h(s;z,w;t)=a(z,w)\partial_z h(s;z,w;t)+b(w)\partial_w h(s;z,w;t),\;\; 0\leq s\leq t.
\end{aligned}
\end{equation}
In the following four subsections, we aim to prove \eqref{partical3} step by step.

\subsection{Calculation of $\frac{\partial h(s;z,w;t)}{\partial t}$}
In this subsection, we calculate $\frac{\partial h(s;z,w;t)}{\partial t}$ based on \eqref{abbr}.
\begin{equation}\label{dev1}
\begin{aligned}
    \frac{\partial h(s;z,w;t)}{\partial t}=&\frac{\partial}{\partial t}A(s;w;t)\left[E(w;t)z+M(w;t)H(s;w;t)\right]\\&
    +A(s;w;t)\left[\frac{\partial}{\partial t}E(w;t)\,z+M(w;t)H(s;w;t)\right]\\&
    +A(s;w;t)\left[E(w;t)z+\frac{\partial}{\partial t}M(w;t)H(s;w;t)\right]\\&
    +A(s;w;t)\left[E(w;t)z+M(w;t)\frac{\partial}{\partial t}H(s;w;t)\right].
\end{aligned}
\end{equation}
Recall the definition of incomplete gamma function in $H(s;w;t)$, namely, $\Gamma(\alpha;x)=\int_x^\infty y^{\alpha-1}\mathrm{e}^{-y}\mathrm{d}y$. Through standard differentiation, we can readily obtain
\begin{equation}
\begin{aligned}
    &\frac{\partial}{\partial t}A(s;w;t)=A(s;w;t)c(w)\delta \mathrm{e}^{\delta(s-t)},\\&
    \frac{\partial}{\partial t}E(w;t)=-uE(w;t),\\&
    \frac{\partial}{\partial t}M(w;t)=-uM(w;t),\\&
    \frac{\partial}{\partial t}H(s;w;t)=-c(w)\delta \mathrm{e}^{\delta(s-t)}X(s;w;t)^{\alpha-1}\mathrm{e}^{-X(s;w;t)}.
\end{aligned}
\end{equation}

\subsection{Calculation of $\frac{\partial h(s;z,w;t)}{\partial z}$}
In this subsection, we calculate $\frac{\partial h(s;z,w;t)}{\partial z}$ based on \eqref{abbr}.
\begin{equation}\label{dev2}
\begin{aligned}
    \frac{\partial h(s;z,w;t)}{\partial z}=&A(s;w;t)E(w;t).
\end{aligned}
\end{equation}

\subsection{Calculation of $\frac{\partial h(s;z,w;t)}{\partial w}$}
In this subsection, we calculate $\frac{\partial h(s;z,w;t)}{\partial w}$ based on \eqref{abbr}.
\begin{equation}\label{dev3}
\begin{aligned}
    \frac{\partial h(s;z,w;t)}{\partial w}=&\frac{\partial}{\partial w}A(s;w;t)\left[E(w;t)z+M(w;t)H(s;w;t)\right]\\&
    +A(s;w;t)\left[\frac{\partial}{\partial w}E(w;t)\,z+M(w;t)H(s;w;t)\right]\\&
    +A(s;w;t)\left[E(w;t)z+\frac{\partial}{\partial w}M(w;t)H(s;w;t)\right]\\&
    +A(s;w;t)\left[E(w;t)z+M(w;t)\frac{\partial}{\partial w}H(s;w;t)\right].
\end{aligned}
\end{equation}
It can be easily calculated that 
\begin{equation}
\begin{aligned}
    &\frac{\partial}{\partial w}A(s;w;t)=-A(s;w;t)\frac{v}{\delta}\mathrm{e}^{\delta(s-t)},\\&
    \frac{\partial}{\partial w}E(w;t)=\frac{v}{\delta}E(w;t),\\&
    \frac{\partial}{\partial w}M(w;t)=\frac{uv}{\delta^2c(w)}M(w;t),\\&
    \frac{\partial}{\partial w}H(s;w;t)=
    \frac{v}{\delta}\mathrm{e}^{\delta(s-t)}X(s;w;t)^{\alpha-1}\mathrm{e}^{-X(s;w;t)}-\frac{v}{\delta}[-c(w)]^{\alpha-1}\mathrm{e}^{c(w)}.
\end{aligned}
\end{equation}

\subsection{The Partial Differential Equation Satisfied by $h(s;z,w;t)$}
Assemble the results in \eqref{dev1}, \eqref{dev2}, and \eqref{dev3}, one checks by direct (term-by-term) cancellation that \eqref{partical3} holds.

\newpage
\section{Hierarchical solver for $\mathcal{B}_{p,q}$ by Layers}
\begin{algorithm}[h!]
\caption{Hierarchical solver for $\mathcal{B}_{p,q}$ by Layers $L=p+q$}
\label{layer}
\begin{algorithmic}[1]
\Require $D,D_1\in\mathbb{R}^{N\times N}$; scalars $u,v,\delta$; max layer $L_{\max}$.
\Ensure All $\{\mathcal{B}_{p,q}\}$ for layers $L=0,\dots,L_{\max}$.
\Function{SolveRight}{$M,\,Y$}
  \Comment Return $X$ such that $X\,M=Y$ (Use LU factorization \cite{MatrixComputations2013})
  \State Factorize $M^\top=L\,U$
  \State Solve $L\,U\,X^\top=Y^\top$ for $X^\top$; \Return $X$
\EndFunction
\State $\mathcal{B}_{0,0}=\bm{I}_N$ is given \Comment Layer $L=0$
\For{$L=1$ \textbf{to} $L_{\max}$} \Comment Advance by layer
  \State \textit{// Previous-layer terms: } $Y_k:=\mathcal{B}_{k,\,L-1-k}$ for $k=0,\dots,L-1$; set $Y_{-1}=Y_L=0$
  \State \textit{// Right-multiplication blocks: } $M_p:=D-\big(up+\delta(L-p)\big)\bm{I}_N$ for $p=0,\dots,L$
  \State \textit{// Same-layer back-substitution (right $\to$ left): let $Z_p:=\mathcal{B}_{p,\,L-p}$}
  \State $Z_L \gets \Call{SolveRight}{M_L,\ -\,Y_{L-1}D_1}$
  \For{$p=L-1$ \textbf{downto} $0$}
    \State $RHS \gets -\,Y_{p-1}D_1 \;-\; v\,p\,Y_p \;-\; v(p+1)\,Z_{p+1}$
    \State $Z_p \gets \Call{SolveRight}{M_p,\ RHS}$
  \EndFor
  \State \textit{// Write back this layer}
  \For{$p=0$ \textbf{to} $L$} \State $\mathcal{B}_{p,\,L-p} \gets Z_p$ \EndFor
\EndFor
\end{algorithmic}
\end{algorithm}

\newpage

\bibliographystyle{siam}
\bibliography{GeneExpression}
\end{document}